\documentclass{llncs}
\usepackage{hyperref}
\usepackage[english]{babel}
\usepackage[utf8]{inputenc}
\usepackage[T1]{fontenc}
\usepackage{amssymb,amsmath}
\usepackage{tikz}
\usepackage{booktabs}
\usepackage{tabularx}
\usepackage{color}
\usepackage{colortbl}
\usepackage[linesnumbered,boxed]{algorithm2e}

\newif\ifanon
\anonfalse

\DeclareMathOperator{\Res}{Res}

\newcommand{\bZ}{\mathbb{Z}}
\newcommand{\bF}{\mathbb{F}}
\newcommand{\bQ}{\mathbb{Q}}
\def\GF#1{\ensuremath{\bF_{#1}}}

\newcommand{\twolinecell}[2][r]{%
\begin{tabular}[#1]{@{}c@{}}#2\end{tabular}}

\title{A kilobit hidden SNFS discrete logarithm computation}

\ifanon
\author{}
\institute{}
\else
\author{Joshua Fried\inst{1}, Pierrick Gaudry\inst{2}, Nadia Heninger\inst{1}, Emmanuel Thom\'e\inst{2}}

\institute{University of Pennsylvania
  \and
  INRIA, CNRS, Université de Lorraine}
\fi
\begin{document}
\maketitle

\begin{abstract}
  We perform a special number field sieve discrete logarithm
  computation in a 1024-bit prime field.  To our knowledge, this is
  the first kilobit-sized discrete logarithm computation ever reported
  for prime fields.  This computation took a little over two months of
  calendar time on an academic cluster using the open-source CADO-NFS
  software.

  Our chosen prime $p$ looks random, and $p-1$ has a 160-bit prime
  factor, in line with recommended parameters for the Digital
  Signature Algorithm.  However, our $p$ has been trapdoored in such a
  way that the special number field sieve can be used to compute discrete
  logarithms in $\mathbb{F}_p^*$, yet detecting that $p$ has this trapdoor
  seems out of reach.  Twenty-five years ago, there was considerable
  controversy around the possibility of backdoored parameters for DSA.
  Our computations show that trapdoored
  primes are entirely feasible with current computing technology.
  We also describe special number field sieve discrete log 
  computations carried out for multiple conspicuously weak primes found in 
  use in the wild.

    As can be expected from a trapdoor mechanism which we say is hard to
    detect, our research did not reveal any trapdoored prime in wide use.
    The only way for a user to defend against a hypothetical trapdoor of
    this kind is to require verifiably random primes.
\end{abstract}

\section{Introduction}

In the early 1990's, NIST published draft standards for what later
became the Digital Signature Algorithm (DSA)~\cite{FIPS-186-4}.  DSA is
now widely used.  At the time, many members of the cryptographic
community voiced concerns about the proposal.  Among these concerns
were that the standard encouraged the use of a global common prime
modulus $p$~\cite{RiHeAnLy92}, and that a malicious party could
specially craft a trapdoored prime so that signatures would be easier
to forge for the trapdoor
owner~\cite{lenstra-report}.
This latter charge was the subject of a remarkable panel at Eurocrypt
1992~\cite{EC:DLLMORS92,cryptolog}. Most of the panelists agreed that it
appeared to be difficult to construct an undetectably trapdoored modulus,
and that such trapdoors appeared unlikely.  To protect against possible
trapdoored primes, the Digital Signature Standard suggests that primes
for DSA be chosen in a ``verifiably random'' way, with a published seed
value~\cite{C:SmiBra92}.  Yet DSA primes used in the wild today are
seldom published with the seed.

Concerns about cryptographic backdoors have a long history (for instance,
it has been formalized as ``kleptography'' in the 90's~\cite{C:YouYun96})
and regained prominence in recent
years since the disclosure of NSA documents leaked by Edward Snowden.  A
set of leaked documents published in September 2013 by the NY
Times~\cite{nytimes:nsa:13}, The Guardian~\cite{guardian:nsa:13}, and
ProPublica~\cite{propublica:nsa:13} describe an NSA ``SIGINT Enabling
Project'' that included among its goals to ``Influence policies,
standards, and specification for commercial public key technologies''.
The newspaper articles describe documents making specific reference to a
backdoor in the Dual EC random number generator, which had been
standardized by both NIST and ANSI.  NIST responded by withdrawing its
recommendation for the Dual EC DRBG, writing ``This algorithm includes
default elliptic curve points for three elliptic curves, the provenance
of which were not described.  Security researchers have highlighted the
importance of generating these elliptic curve points in a trustworthy
way.  This issue was identified during the development process, and the
concern was initially addressed by including specifications for
generating different points than the default values that were provided.
However, recent community commentary has called into question the
trustworthiness of these default elliptic curve
points''~\cite{nist-bulletin}.  There is evidence that the ability to
backdoor the Dual EC algorithm has been exploited in the wild: Juniper
Networks had implemented Dual EC in NetScreen VPN routers, but had used
it with custom-generated parameters.  In December 2015 Juniper published
a security advisory~\cite{juniper-announcement} announcing that an
attacker had made unauthorized modifications to the source code for these
products to substitute a different curve point in the Dual EC
implementation~\cite{CCS:CMGFCG16}.

In this paper, we demonstrate that constructing and exploiting trapdoored
primes for Diffie-Hellman and DSA is feasible for 1024-bit keys with
modern academic computing resources.  Current estimates for 1024-bit
discrete log in general suggest that such computations are likely within
range for an adversary who can afford hundreds of millions of dollars of
special-purpose hardware~\cite{Logjam}.  In contrast, we were able to
perform a discrete log computation on a specially trapdoored prime in two
months on an academic cluster.  While the Dual EC algorithm appears to
have only rarely been used in practice~\cite{dualec-usenix}, finite-field
Diffie-Hellman and DSA are cornerstones of public-key cryptography. We
neither show nor claim that trapdoored primes are currently in
use.  However, the near-universal
failure of implementers to use verifiable prime generation practices
means that use of weak primes would be undetectable in practice and
unlikely to raise eyebrows.

\subsection*{The special number field sieve trapdoor}

The Number Field Sieve (NFS), still very much in its infancy at the
beginning of the 1990's, was originally proposed as an integer
factoring algorithm~\cite{LeLe93}. Gordon adapted the algorithm to
compute discrete logarithms in prime fields~\cite{Gordon93b}. Both for
the integer factoring and the discrete logarithm variants, several
theoretical and computational obstacles had to be overcome before the
NFS was practical to use for large scale computations. For the past
twenty years, the NFS has been routinely used in record computations,
and the underlying algorithms have been thoroughly improved. The NFS
is now a versatile algorithm which can handle an arbitrary prime $p$,
and compute discrete logarithms in $\GF p^*$ in asymptotic time
$L_p(1/3,(64/9)^{1/3})^{1+o(1)}$, using the usual $L$-notation
(see~§\ref{sec:l-notation}).

Current computational records for the number field sieve include a 768-bit 
factorization of an RSA modulus, completed in December 2009 by Kleinjung et al.~\cite{RSA768} 
and a 768-bit discrete log for a safe prime, completed in June 2016 by Kleinjung et al.~\cite{DLP768}.

Very early on in the development of NFS, it was observed that the
algorithm was particularly efficient for inputs of a special form.
Some composite integers are particularly amenable to being factored by
NFS, and primes of a special form allow easier computation of discrete
logarithms. This relatively rare
set of inputs defines the Special Number Field
Sieve (SNFS).  It is straightforward to start with parameters that
give a good running time for the NFS---more precisely, a pair of
irreducible integer polynomials meeting certain degree and size
constraints---and derive an integer to be factored, or a prime modulus
for a discrete logarithm.  In general, moving in the other direction,
from a computational target to SNFS parameters, is known to be
possible only in rare cases (e.g.\ the Cunningham project). The
complexity of SNFS is $L_p(1/3,(32/9)^{1/3})^{1+o(1)}$, much less than
its general counterpart.  A 1039-bit SNFS factorization was completed
in 2007 by Aoki et al.~\cite{Aoki2007}.

In 1992, Gordon~\cite{C:Gordon92} suggested several methods that are
still the best known for trapdooring primes to give the best running
time for the SNFS, without the trapdoored SNFS property being
conspicuous\footnote{In 1991, another method was suggested by Lenstra
in~\cite{lenstra-report}, and played a role in triggering the Eurocrypt
panel~\cite{EC:DLLMORS92}. Gordon's trap design is more general.}.  Most of his analysis remains valid, but there has been
significant improvement in the NFS algorithm in the past 25 years.  In
the early days, an NFS computation had to face issues of dealing with
class groups and explicit units. That meant much less flexibility in
creating the trapdoor, to the point that it was indeed difficult to
conceal it. It is now well understood that these concerns were artificial
and can be worked around~\cite{Schirokauer93}, much to the benefit of the
trapdoor designer. Gordon's analysis
and much of the discussion of trapdoored DSA primes in 1992 focused on 512-bit
primes, the suggested parameter sizes for NIST's DSS draft at the
time.  However, 25 years later, 1024-bit targets are of greater
cryptanalytic interest.  

We update the state of the art in crafting trapdoored 1024-bit primes for
which practical computation of discrete logarithms is possible, and demonstrate
that exploitation is practical by performing a 1024-bit SNFS discrete log computation.

We begin by reviewing in Section~\ref{sec:the-wild-wild-west} the
origin of primes found in multiple practical cryptographic contexts.
Section~\ref{sec:nfs} recalls a brief background on the Number Field
Sieve.
In Section~\ref{sec:heidi}, we reevaluate Gordon's work on trapdooring primes for SNFS given the
modern understanding of the algorithm, and explain for a given target size which
        polynomial pair yields the fastest running time. This answers a
        practical question for the problem at hand---how to optimally select trapdoor parameters
        to simplify computations with the prime---and is also of wider
        interest for NFS-related computations.
        
We then run a full 1024-bit experiment to show that trapdoored primes are indeed a
        practical threat, and perform a full 1024-bit SNFS discrete log computation for our trapdoored prime.
        We describe our computation in Section~\ref{sec:computation}.
We show how various adaptations to the block Wiedemann
        algorithm are essential to minimizing its running time, compared
        to previous computations of the same kind.
We detail the descent procedure, and the various challenges
        which must be overcome so as to complete individual logs in a short time.
We also provide an extensive appendix giving details on the analysis of individual logarithm computation 
in Appendix~\ref{appendix:individual}, as this portion of the computation is not well detailed in the literature.

Finally, we evaluate the impact of our results in Section~\ref{sec:discussion}.  Our computation required roughly a factor of 10 less resources than
the recent 768-bit GNFS discrete log announced by Kleinjung et al.  However, we have found a number of primes
amenable to non-hidden SNFS DLP computations in use in the wild.  We
describe additional SNFS computations we performed on these
primes in Section~\ref{sec:nonhidden}.

\section{Modern security practices for discrete log cryptosystems}
\label{sec:the-wild-wild-west}

\paragraph{Verifiable prime generation.} It is legitimate to wonder whether one should worry about trapdoored
primes at all. Good cryptographic practice recommends that publicly
agreed parameters must be ``verifiably random''. For example, appendix
A.1.1.2 of the FIPS 186 standard~\cite{FIPS-186-4} proposes a method
to generate DSA primes~$p$ and~$q$ from a random seed and a hash
function, and suggests that one should publish that seed alongside
with $p$ and $q$. The publication of this seed is marked
\emph{optional}. Primes of this type are widely used for a variety of
cryptographic primitives; for example NIST SP 800-56A specifies that
finite-field parameters for key exchange should be generated using
FIPS 186~\cite[§5.5.1.1]{sp800-56a}.

While it is true that some standardized
cryptographic data includes ``verifiable
randomness''\footnote{This still leaves the question of whether the
  seed is honest, see e.g.~\cite{scott99sci.crypt,bada55}. We do not
  address this concern here.} or rigidity derived from ``nothing up my sleeve'' numbers, it is noteworthy that this is not
always the case. For example, both France and China standardized
elliptic curves for public use without providing any sort of
justification for the chosen parameters~\cite[§3.1]{bada55}.
RFC
5114~\cite{rfc5114} specifies a number of groups for use with
Diffie-Hellman, and states that the parameters were drawn from NIST
test data, but neither the NIST test data~\cite{80056aex} nor RFC 5114 itself contain
the seeds used to generate the finite field parameters.  In a similar
vein, the origin of the default 2048-bit prime
in the Helios voting system used in the most recent IACR Board of Directors
Election in 2015 is undocumented.  Most users would have to go out of their
way to generate verifiable primes: the default behavior of OpenSSL does
not print out seeds when generating Diffie-Hellman or DSA
parameter sets.  
However, some implementations do provide seeds. Java's
\texttt{sun.security.provider} package
specifies hard-coded 512-, 768-, and 1024-bit groups together with the FIPS 186
seeds used to generate them.

\paragraph{Standardized and hard-coded primes.}
It is also legitimate to wonder whether one should be concerned about widespread 
reuse of primes.  For modern computers, prime generation is much less computationally burdensome than 
in the 1990s, and any user worried about a backdoor could easily generate their own group parameters.
However, even today, many applications use standardized or hard-coded primes for
Diffie-Hellman and DSA. We illustrate this by several examples.

In the TLS protocol, the server specifies the group parameters that
the client and server use for Diffie-Hellman key exchange.  Adrian et
al.~\cite{Logjam} observed in 2015 that 37\% of the Alexa Top 1
Million web sites supported a single 1024-bit group for Diffie-Hellman
key exchange.  The group parameters were hard-coded into Apache 2.2,
without any specified seed for verification.  They also observed that
in May 2015, 56\% of HTTPS hosts selected one of the 10 most common
1024-bit groups when negotiating ephemeral Diffie-Hellman key
exchange.  Among 13 million recorded TLS handshakes negotiating
ephemeral Diffie-Hellman key exchange, only 68,000 distinct prime
moduli were used.  The TLS 1.3 draft restricts finite-field
Diffie-Hellman to a set of five groups modulo safe primes ranging in
size from 2048 to 8196 bits derived from the nothing-up-my-sleeve
number $e$~\cite{tls13-negotiatedff}.

In the IKE protocol for IPsec, the initiator and responder negotiate a group for
Diffie-Hellman key exchange from a set list of pre-defined groups;
Adrian et al.\ observed that 66\% of IKE responder hosts preferred the 1024-bit
Oakley Group 2 over other choices. The Oakley groups specify a collection of primes
derived from a ``nothing-up-my-sleeve'' number, the binary expansion
of $\pi$, and have been built into standards, including IKE and SSH, for decades~\cite{rfc2412}.
        The additional finite-field Diffie-Hellman groups specified
            in RFC~5114
are widely used in practice: Internet-wide scans from September
2016 found that over 900,000 (2.25\%) of TLS hosts on port 443 chose
these groups~\cite{CCS:DAMBH15}.
Scans from February 2016 of IKE hosts on port 500 revealed that 340,000
(13\%) supported the RFC 5114
finite-field Diffie-Hellman parameters~\cite{subgroup}.

RFC 4253 specifies two groups that must be supported for SSH Diffie-Hellman key exchange:
Oakley Group 2 (which is referred to as SSH group 1) and Oakley Group 14, a 2048-bit prime.  SSH group 1 key exchange was disabled by default in OpenSSH version 7.0, released in August 2015~\cite{openssh7}.
Optionally, SSH clients and servers may negotiate a different group using the group exchange handshake.  However, OpenSSH
chooses the group negotiated during this exchange from a pre-generated list that is generally shipped with the software package.
The \texttt{/etc/ssh/moduli} file on an Ubuntu 16.04 machine in our cluster contained 267 entries in size between 1535 and 8191 bits.  
The 40 to 50 primes at each
size appear to have been generated by listing successive primes from a fixed starting point, and differ only in the least significant handful of bits.
We examined data from a full IPv4 SSH scan performed in October 2015~\cite{subgroup} that offered Diffie-Hellman group exchange only,
and found 11,658 primes in use from 10.9 million responses, many of which could be clustered into groups differing only in a handful of least significant bits.

The SSH protocol also allows servers to use long term DSA keys to authenticate themselves to clients.
We conducted a scan of a random 1\% portion of the IPv4 space for hosts running SSH servers on port 22 with 
DSA host keys in September 2016, and found that most hosts seemed to generate unique primes for
their DSA public keys. The scan yielded 27,380 unique DSA host keys from 32,111 host servers, of which only 557 shared a prime with another key.  DSA host key authentication was also disabled by default in OpenSSH 7.0~\cite{openssh7}.

\paragraph{1024-bit primes in modern cryptographic deployments.}
It is well understood that 1024-bit factorization and discrete log
computations are within the range of government-level adversaries~\cite{Logjam}, but
such computations are widely believed by practitioners to be
\emph{only} within the range of such adversaries, and thus that these
key sizes are still safe for use in many cases.  While NIST has
recommended a minimum prime size of 2048 bits since 2010~\cite{sp800-131a}, 1024-bit
primes remain extremely common in practice.  Some of this is due to
implementation and compatibility issues.  For example, versions of
Java prior to Java 8, released in 2014, did not support Diffie-Hellman
or DSA group sizes larger than 1024 bits.  DNSSEC limits DSA keys to a
maximum size of 1024-bit keys~\cite{rfc6781}, and stated, in 2012,
that with respect to RSA keys, ``To date, despite huge efforts, no one has broken
a regular 1024-bit key; \dots it is estimated that most zones can
safely use 1024-bit keys for at least the next ten years.''  SSL Labs
SSL Pulse estimated in September 2016 that 22\% of the 200,000 most
popular HTTPS web sites performed a key exchange with 1024-bit
strength~\cite{ssllabs}.

\section{The Number Field Sieve for discrete logarithms}
\label{sec:nfs}

\subsection{The NFS setting}
We briefly recall the Number Field Sieve (NFS) algorithm for computing
discrete logarithms in finite fields. This background is classical and
can be found in a variety of references. NFS appeared first as an algorithm for
factoring integers~\cite{LeLe93}, and has been adapted to the computation of
discrete logarithms over several works~\cite{Gordon93b,JoLe03}.

Let $\GF p$ be a prime field, let $\gamma\in\GF p^*$ be an element of prime
order $q \mid p-1$. We wish to solve discrete logarithms in $\langle
\gamma\rangle$.
The basic outline of the NFS-DL algorithm is as follows:

\begin{description}
    \item[Polynomial selection.] Select irreducible integer
        polynomials $f$ and $g$, sharing a common root $m$ modulo $p$.
        Both polynomials define number fields, which we denote by
        $\bQ(\alpha)$ and $\bQ(\beta)$.
    \item[Sieving.] Find many pairs $a,b$ such that the two
        integers (called {\em norms} -- albeit improperly if $f$ or $g$
        are not monic) 
        $\Res(f(x),a-bx)$ and $\Res(g(x),a-bx)$ factor completely into
        primes below a chosen smoothness bound $B$.
    \item[Filtering.] Form multiplicative combinations of the
        $(a,b)$ pairs to reduce the number of prime ideals appearing in the
        corresponding ideal factorizations.
    \item[Compute maps.] Compute $q$-adic characters
        (known as Schirokauer maps~\cite{Schirokauer93}).
        This yields a relation matrix which can be
        written as $(M\|S)$, with $M$ the block with ideal
        valuations, and $S$ the block with Schirokauer maps.
    \item[Linear algebra.] Solve the linear system
        $(M\|S)x=0$, which gives \emph{virtual logarithms}.
    \item[Individual logarithm.] Given a target value $z\in\langle
        \gamma\rangle$, derive its logarithm as a linear combination of
        a subset of the virtual logarithms.
\end{description}

\subsection{Complexity analysis}

\label{sec:l-notation}
NFS complexity analysis involves the usual $L$-notation,
defined as
\begin{equation}
L_p(e,c)=\exp(c(\log p)^e(\log\log p)^{1-e}).
\end{equation}
This notation interpolates between polynomial
($e=0$) and exponential ($e=1$) complexities. It adapts well to
working with smoothness probabilities. In this formula and elsewhere
in the paper, we use the $\log$ notation for the natural
logarithm. In the few places where we have formulae that involve
bit sizes, we always use $\log_2$ for the logarithm in base 2.

The polynomials~$f$ and~$g$ have a crucial impact on the size of the
integers $\Res(f(x),a-bx)$ and $\Res(g(x),a-bx)$, and therefore on the
probability that these integers factor into primes below $B$ (in other
words, are \emph{$B$-smooth}).

\begin{table}
    \begin{center}
      \begin{tabular}{l @{\hspace{2em}} c @{\hspace{2em}} c @{\hspace{2em}} c @{\hspace{1em}} c @{\hspace{1em}} c}
        \toprule
            Variant &
            $\deg f$ & $\|f\|$ &
            $\deg g$ & $\|g\|$ &
            \twolinecell{complexity\\exponent}\\
            \midrule
            General NFS (base-$m$) & $d$ & $p^{1/(d+1)}$ & $1$ &
            $p^{1/(d+1)}$ & $(64/9)^{1/3}$\\
            General NFS (Joux-Lercier) & $d'+1$ & $O(1)$ & $d'$ &
            $p^{1/(d'+1)}$ & $(64/9)^{1/3}$\\
            Special NFS (for example) & $d$ & $O(1)$ & $1$ &
            $p^{1/(d+1)}$ & $(32/9)^{1/3}$\\
            \bottomrule
        \end{tabular}
      \caption{\label{table:polyselect}
        Polynomial selection choices for
        NFS variants.}
    \end{center}
\end{table}

The analysis of the NFS depends on the prime $p$. When no assumptions are made
on $p$, we have the so-called \emph{general number field
sieve} (GNFS). It is possible to perform polynomial selection so that
$(\deg f, \deg g)$ is $(1,d)$ (by choosing $m \approx N^{1/d}$ and writing $p$ in ``base-$m$'', or similar),
or $(d'+1,d')$ (the Joux-Lercier method~\cite{JoLe03}).  The degrees
$d$ and $d'$ in each case are
integer parameters, for which an educated guess is provided by their
asymptotic optimum, namely
$d=\left(3{\log p}/{\log\log p}\right)^{1/3}$, and $d'=d/2$.
Both approaches lead to an overall complexity
$L_p(1/3,(64/9)^{1/3})^{1+o(1)}$ for a discrete logarithm computation, as indicated in
Table~\ref{table:polyselect}, where $\|f\|$ denotes the maximum absolute value of
the coefficients of $f$.

In contrast, some prime numbers are such that \emph{there exist}
exceptionally small polynomial pairs $(f,g)$ sharing a common root
modulo $p$. This makes a considerable difference in the efficiency of
the algorithm, to the point that the exponent in the complexity drops
from $(64/9)^{1/3}$ to $(32/9)^{1/3}$---a difference which is also
considerable in practice.  In most previously considered cases, this
special structure is clear from the number itself.  For the SNFS
factorization performed by Aoki et al.\ for the composite integer
$2^{1039}-1$, they chose $f(x) = 2x^6-1$ and $g(x) =
x-2^{173}$~\cite{Aoki2007}.

\section{Heidi hides her polynomials}
\label{sec:heidi}

Early on in the development of NFS, Gordon~\cite{C:Gordon92} suggested
that one could craft primes so that SNFS polynomials exist, but may
not be apparent to the casual observer.  Heidi, a mischievous designer
for a crypto standard, would select a pair of SNFS polynomials to her
liking \emph{first}, and publish only their resultant $p$ (if it is
prime) \emph{afterwards}. The hidden trapdoor then consists in the
pair of polynomials which Heidi used to generate $p$, and that she can
use to considerably ease the computation of discrete logarithms in \GF
p.

Twenty-five years later, we reconsider the best-case scenario for
Heidi: given a target size, what type of polynomial pair will give the
fastest running time for a discrete logarithm computation?  For the
current state of the art in algorithmic development and computation
power, is there a parameter setting for which the computations are
simultaneously within reach, Heidi can efficiently generate a
trapdoored prime, and defeat attempts at unveiling it?

\subsection{Best form for SNFS polynomials}

The Special Number Field Sieve has been mostly used in the context of
integer factorization, in particular for numbers from the Cunningham
Project. In that case the integers are given, and typically the only
way to find SNFS polynomials for these numbers is to take one linear
polynomial with large coefficients and a polynomial of larger degree,
with tiny coefficients. In our situation, the construction can go the
opposite way: we are free to choose first the form of the polynomials,
hoping that their resultant will be a prime number.  We thus have more
freedom in the construction of our polynomial pair.

Let $n$ be the number of bits of the SNFS prime $p$ we will construct.
We consider first the case where we have two polynomials $f$ and $g$ that
are non-skewed, i.e. all the coefficients of each polynomial have roughly
the same size.
We denote $d_f$ and $d_g$ their respective degrees, and
$\|f\|$ and $\|g\|$ the respective maximum absolute values of their
coefficients. Since the resultant must be almost equal to $p$, we have
\begin{equation}\label{eq:resultant}
d_f\log_2\|g\| + d_g\log_2\|f\| \approx n.
\end{equation}
Let $A$ be a bound on the $a$ and $b$ integers we are going to consider
during relation collection. Then the product of the norms that have to be
tested for smoothness can be approximated by $\|f\|\, \|g\|\, A^{d_f+d_g}$.
We will try to make its size as small as possible, so we want to
minimize
\begin{equation}\label{eq:lognorm}
\log_2\|f\| + \log_2\|g\| + (d_f+d_g)\log_2 A.
\end{equation}
Of course, the value taken by $A$ will also depend on the size of
the norms.  If this size is larger than expected, then the
probability of finding a relation is too small and the sieving range
corresponding to $A$ will not allow the creation of enough relations. But
assuming $A$ is fixed is enough to compare various types of polynomial
constructions: if one of them gives larger norms, then for this
construction, the value of $A$ should be larger, leading to even larger norms. In other
words, the optimal value is unchanged whether we consider $A$ fixed or
let it depend on $d_f$ and $d_g$.

\subsubsection{The best asymmetric construction is the classical SNFS.}
We first analyze the case where $d_f$ and $d_g$ are distinct. Let us
assume $d_f>d_g$. We first remark that subject to
constraint~\eqref{eq:resultant}, Expression~\eqref{eq:lognorm} is minimized
by taking $\|f\|$ as small as possible (i.e.\ $\log_2\|f\|=0$) and
$\log_2\|g\|=n/d_f$. This yields an optimal
norm size equal to $n/d_f+(d_f+d_g)\log_2 A$. It follows that given $d_f$,
we should choose $d_g$ to be minimal, which leads us precisely to
the classical case, the example construction listed in the third row of
Table~\ref{table:polyselect}. The optimal $d_f$ yields an optimal norm
size equal to
$2\sqrt{n\log_2 A}$.

\subsubsection{An all-balanced construction.}
In many situations, the
optimal value is obtained by balancing each quantity as much as possible.
Unfortunately, this is suboptimal in our case. If 
$d_f=d_g$, Expression~\eqref{eq:lognorm} becomes $n/d_f+2d_f\log_2 A$.
Choosing the best
possible value for $d_f$, we obtain $2\sqrt{2n\log_2 A}$. This is much
worse than in the classical construction and in fact, pushing the
analysis to its end would lead to a GNFS complexity with a $(64/9)^{1/3}$
exponent.

\subsubsection{More general constructions.} Unfortunately, it seems to be
impossible to combine the SNFS construction with Coppersmith's multiple
number field strategy~\cite{JC:Coppersmith93,Matyukhin03,CoSe06} and obtain a complexity with an exponent smaller
than $(32/9)^{1/3}$. Any linear combination of $f$ and $g$ will lead to a
polynomial having both high degree and large coefficients, which must be
avoided to achieve SNFS complexity.

In principle, one could also perform an analysis allowing skewed
polynomials, where the ratio between two consecutive coefficients is
roughly a constant different from 1. This general analysis would require
still more parameters than the one we did, so we skip the details, since
we did not find a situation where this could lead to a good asymptotic
complexity.

\subsection{Hiding the special form}

The conclusion of the previous discussion is that the best form for a
pair of SNFS polynomials is still the same as the one considered by
Gordon more than 20 years ago. His discussion about how to hide them is
still valid.  We recall it here for completeness.

The goal is to find a prime $p$, and possibly a factor $q$ of $p-1$,
together with an SNFS pair of polynomials, such that from the
knowledge of $p$ (and $q$) it is harder to guess the SNFS pair of
polynomials than to run a discrete logarithm computation with the
general NFS algorithm, or using Pollard Rho in the subgroup of order
$q$.

We enumerate requirements on the construction below.

\subsubsection{The polynomial $f$ must be chosen within a large enough set.}

If $f$ is known, then its roots modulo $p$ can be computed.  With the
Extended Euclidean Algorithm, it can be efficiently checked whether
one of them is equal to a small rational number modulo $p$. If this is
the case, then the numerator and the denominator are (up to sign) the
coefficients of $g$. Therefore, if $f$ has been chosen among a small
set, an exhaustive search over the roots modulo $p$ of all these
polynomials will reveal the hidden SNFS pair of polynomials.  Thus we
must choose $f$ from a large enough set so that this exhaustive search
takes at least as much time as a direct discrete logarithm
computation.

\subsubsection{The two coefficients of $g$ must be large.}

If $g$ is a monic polynomial $g=x-g_0$, then, since $p=f(g_0)$, the
most significant bits of $p$ depend only on $g_0$ and the leading
coefficient $f_d$ of $f$. In that case, recovering the hidden
SNFS polynomials reduces to an exhaustive search on the leading
coefficient of~$f$: we can use the LLL algorithm to minimize the other
coefficients of $f$ by writing a multiple of $p$ as a sum of powers of
$g_0$.
Examining the least significant bits of $p$ shows
that having a polynomial $g$ with a constant term equal to $1$ is equally
bad. More generally, having one of the coefficients of $g$ belonging to a
small set also leads to a faster exhaustive search than if both are
large. In the following, we will therefore always consider linear
polynomials $g$ for which the two coefficients have similar sizes;
compared to using a monic $g$, this has only a marginal impact on the
effectiveness of the SNFS efficiency in our context.

\subsubsection{Attempts to unveil the trapdoor}
\label{sec:unveil}

Heidi does not want her trapdoor to be unveiled, as she would not be able
to plausibly deny wrongdoing. It is therefore highly important that Heidi
convinces herself that the criteria above are sufficient for the trapdoor
to be well hidden. We tried to improve on the method mentioned above that
adapts to monic $g$. In particular, we tried to take advantage of
the possibility that the leading coefficient of $f$ might be divisible by small primes.
This did not lead to a better method.

\subsection{Adapting the prime to the hider's needs}

\subsubsection{Algorithm to build a DSA-like prime.}

In Algorithm~\ref{alg:gordon}, we recall the method of Gordon to
construct hidden SNFS parameters in a DSA setting. The general idea is to
start from the polynomial $f$ and the prime $q$, then derive a polynomial
$g$ such that $q$ divides the resultant of $f$ and $g$ minus 1, and only
at the end check if this resultant is a prime $p$. This avoids the costly
factoring of $p-1$ that would be needed to check whether there is a
factor of appropriate size to play the role of $q$.  Our version is
slightly more general than Gordon's, since we allow signed coefficients for the
polynomials. As a consequence, we do not ensure the sign of the
resultant, so that the condition $q\mid p-1$ can fail. This explains the
additional check in Step~8. The size of the coefficients of $f$ are also
adjusted so that an exhaustive search on all the polynomials will take
more or less the same time as the Pollard Rho algorithm in the subgroup
of order $q$, namely $2^{s_q/2}$ where $s_q$ is the bit-length of $q$.

In Step~6, it is implicitly assumed that $2^{s_q}$ is smaller than
$2^{s_p/d}/\|f\|$, that is $s_q < s_p/d - s_q/2(d+1)$. This condition
will be further discussed in the next subsection. We note however that if
it fails to hold by only a few bits, it is possible to run the algorithm
and hope that the root $r$ produced at Step~5 will be small enough. We
can expect that $r$ will behave like a uniformly random element modulo
$q$, so that the probability that this event occurs can be estimated.

\SetAlCapSkip{1ex}
\begin{algorithm}
\SetKwInOut{Input}{Input}\SetKwInOut{Output}{Output}
\Input{The bit-sizes $s_p$ and $s_q$ for $p$ and $q$; the degree $d$ of $f$.}
\Output{HSNFS parameters $f$, $g$, $p$, $q$.}
\BlankLine
Pick a random irreducible polynomial $f$, with $\|f\|\approx 2^{s_q/2(d+1)}$\;
Pick a random prime $q$ of $s_q$ bits\;
Pick a random integer $g_0\approx 2^{s_p/d}/\|f\|$\;
Consider the polynomial $G_1(g_1)=\Res_x(f(x), g_1x+g_0)-1$ of degree $d$
in $g_1$\;
Pick a root $r$ of $G_1$ modulo $q$; if none exists go back to Step~1\;
Add a random multiple of $q$ to $r$ to get an integer $g_1$ of size
$\approx 2^{s_p/d}/\|f\|$\;
Let $p=|\Res_x(f(x), g_1x+g_0)|$\;
If $p$ has not exactly $s_p$ bits or if $p$ is not prime or if $q$ does
not divide $p-1$, then go back to Step~1\;
Return $f$, $g$, $p$, $q$.
\caption{Gordon's hidden SNFS construction algorithm}
\label{alg:gordon}
\end{algorithm}

\subsubsection{Selecting good $f$-polynomials.}

In Algorithm~\ref{alg:gordon}, in the case of failure at Step~5 or
Step~8, we could restart only at Step~2, in order to keep using the same
$f$-polynomial for a while. More generally, the polynomial $f$ could be
given as input of the algorithm, opening the opportunity for the hider
to use a polynomial~$f$ with nice algebraic properties that accelerate the
NFS algorithm. The so-called Murphy-$\alpha$ value~\cite[§3.2]{Murphy99} has a measurable
influence on the probability of the norms to be smooth. A norm of $s$ bits
is expected to have a smoothness probability similar to the one of a random
integer of $s+\frac{\alpha}{\log 2}$ bits. A negative $\alpha$-value is
therefore helping the relation collection.

Experimentally, for an irreducible polynomial of fixed degree over $\bZ$
with coefficients uniformly distributed in an interval, the
$\alpha$-value follows a centered normal law with standard deviation
around $0.94$ (measured empirically for degree $6$).
From this, it is possible to estimate the expected minimum $\alpha$-value
after trying $N$ polynomials: we get $\alpha_{\mathrm{min}} \sim -0.94\sqrt{2\log N}$.

In a set of $2^{80}$ candidates for the $f$-polynomial, we can therefore
expect to find one with an $\alpha$-value around $-10$. But it is {\em a
priori} very hard to find this polynomial, and if it were easy, then it
would not be a good idea for the hider to choose it, because then it would
not be hidden anymore. A compromise is for the hider to try a small
proportion of the candidates and keep the one with the best $\alpha$.
Since checking the $\alpha$-value of a polynomial is not really faster
than checking its roots modulo $p$, the attacker gains no advantage by
knowing that $f$ has a smaller value than average. For instance, after
trying $2^{20}$ polynomials, one can expect to find an $f$ that has an
$\alpha$-value of $-5$ which gives a nice speed-up for the NFS without
compromising the hidden property.

Apart from the $\alpha$-value, another well-known feature of polynomials
that influences the smoothness properties is the number of real roots:
more real roots translates into finding relations more easily. We did not
take this into account in our proof of concept experiments, but this
could certainly also be used as a criterion to select $f$.

\subsection{Size considerations}

Algorithm~\ref{alg:gordon} does not work if the size $s_q$ of the
subgroup order $q$ is too large compared to the size of the coefficients
of the $g$-polynomials that are optimal for the size of $p$. The
condition is
$$s_q < s_p/d - s_q/2(d+1),$$
where $d$ is the optimal degree of the $f$-polynomial for running SNFS on
a prime of $s_p$ bits. We can plug in the asymptotic formula for $d$ in
terms of $s_p$: it is proportional to $(s_p/\log(s_p))^{1/3}$, leading
to a condition of the form
$$ s_q < c(s_p)^{2/3}(\log(s_p))^{1/3} = \log(L_p(2/3, c)),$$
for a constant $c$. Now, $s_q$ will be chosen so that the running time of
Pollard Rho in the subgroup of order $q$ matches the running time of the
NFS algorithm modulo~$p$. The former grows like $2^{s_q/2}$, while the
latter grows like $L_p(1/3, c')\approx 2^{s_p^{1/3}}$. Therefore,
asymptotically, it makes sense to have $s_q$ close to proportional to $s_p^{1/3}$,
and the condition for Algorithm~\ref{alg:gordon} to work is easily
satisfied.

Back in 1992, when Gordon studied the question, the complexity analysis
of the Number Field Sieve was not as well understood, and the available
computing power was far less than today. At that time, $s_q=160$ and
$s_p=512$ were the proposed parameter sizes for DSA,
leading to difficulties satisfying the condition of
Algorithm~\ref{alg:gordon} unless a suboptimal $d$ was chosen. Nowadays,
popular DSA parameters are $s_p=1024$ and $s_q=160$, leaving much room
for the condition to hold, and it is possible to choose $d=6$, which is optimal for our NFS
implementation. Therefore, the relevant parameters for today are
beyond the point where Gordon's algorithm would need to be run with
suboptimal parameters.

\subsection{Reassessing the hiding problem}

The prudent conclusions of cryptographers in the 1990's was that it
might be difficult to put a useful and hard to detect trapdoor in a
DSA prime.  For example in~\cite{EC:DLLMORS92}, Lenstra concludes that
``this kind of trap can be detected'', based on the trap design
from~\cite{lenstra-report}. It is true that whether for Lenstra's trap
method in~\cite{lenstra-report}, or Gordon's trap in
Algorithm~\ref{alg:gordon},
$f$ had to be chosen within a
too-small set given the state of the art with NFS back in
1992.
This stance is also found in reference books from the
time, such as the \emph{Handbook of Applied Cryptography} by Menezes, van Oorschot, and Vanstone~\cite[note §8.4]{MenVanVan97} which remain influential today.

This is
no longer true.
It is now clearly possible to hide an SNFS pair of polynomials for a DSA
prime $p$ of 1024 bits with a 160-bit subgroup. It remains to show that
this SNFS computation is indeed feasible, even with moderate academic
computing resources.

\section{Computation of a 1024-bit SNFS DLP}
\label{sec:computation}

In addition to the computational details, we describe the algorithmic
improvements and parameter choices that played a key role in the
computation.

\definecolor{lightgray}{rgb}{0.9,0.9,0.9}
\begin{table}
    \begin{center}
\setlength{\extrarowheight}{.76ex}
      \begin{tabular*}{\textwidth}{@{\extracolsep{\stretch{1}}}l >{\columncolor{lightgray}}rrrr>{\columncolor{lightgray}}r@{}}
\noalign{\global\belowrulesep=0em\global\aboverulesep=0em}
\toprule
    & sieving\hfil & \multicolumn{3}{c}{linear algebra} & individual log \\
 & & sequence & generator & solution &\\
 \midrule
    cores  & $\approx$3000 & 2056 & 576 & 2056 & 500--352\\
    CPU time (1 core) &
240 years & 123 years & 13 years & 9 years & 10 days \\
\cmidrule(r){3-5}
{calendar time} &
1 month & \multicolumn{3}{c}{1 month} & 80 minutes \\
\bottomrule
\end{tabular*}
    \end{center}
\caption{Our 1024-bit hidden SNFS discrete log computation took around two months of calendar time to complete.  We used a variety of resources for sieving, so the total number of cores in use varied over time.
}
\end{table}

\subsection{Selecting a target}

We ran Algorithm~\ref{alg:gordon} to find a hidden SNFS prime $p$ of
1024 bits such that $\bF_p$ has a subgroup of prime order $q$ of 160
bits. For these parameters, a polynomial $f$ of degree $d=6$ is the most
appropriate. After a small search among the polynomials with (signed)
coefficients of up to 11 bits, we selected
$$ f = 1155\,x^6 + 1090\,x^5 + 440\,x^4 + 531\,x^3 - 348\,x^2 - 223\,x
- 1385,$$ for which the $\alpha$-value is about $-5.0$. The set of all
polynomials with this degree that satisfy the coefficient bound is a
bit larger than $2^{80}$, which is the expected cost of Pollard Rho
modulo $q$.  We note that the cost of testing a polynomial $f$ (a root
finding modulo $p$ and the rational reconstruction of these roots) is
much higher than one step of Pollard Rho (one multiplication modulo
$p$), so this is a conservative setting.

We then ran the rest of Algorithm~\ref{alg:gordon} exactly as it is described.
The resulting public parameters are
{\small
$$
\begin{array}{rcl}
    p & = & 163323987240443679101402070093049155030989439806917519173580070791569\\
    &&  227728932850358498862854399351423733697660534800194492724828721314980\\
    &&  248259450358792069235991826588944200440687094136669506349093691768902\\
    &&  440555341493237296555254247379422702221515929837629813600812082006124\\
    &&  038089463610239236157651252180491
    \\
    q & = & 1120320311183071261988433674300182306029096710473\ , \\
\end{array}
$$
}
and the trapdoor polynomial pair is
{\small
$$
\begin{array}{rcl}
    f & = & 1155\,x^6 + 1090\,x^5 + 440\,x^4 + 531\,x^3 - 348\,x^2 - 223\,x - 1385 \\
    g & = & 567162312818120432489991568785626986771201829237408\,x \\
    && -{} 663612177378148694314176730818181556491705934826717\ . \\
\end{array}
$$}

This computation took 12 core-hours, mostly spent in selecting a
polynomial $f$ with a good $\alpha$-value. No effort was made to optimize
this step.

\subsection{Choosing parameters for the sieving step}

The sieving step (also known as relation collection) consists of finding many
$(a,b)$-pairs such that the two norms $\Res(f(x),a-bx)$ and
$\Res(g(x),a-bx)$ are
simultaneously smooth.

We use the special-$q$ sieving strategy, where we concentrate the search
in the positions where we know in advance that one of the two norms will be
divisible by a large prime: the special $q$. For the
general number field sieve, it is always the case that one norm is much
larger than the other, so it makes sense to choose the special $q$ on the
corresponding side. In our case, the norms have almost the same
size (about 200 bits each), so there is no obvious choice. Therefore, we
decided to sieve with special $q$'s on both sides. As a consequence, the
largest special $q$ that we had to consider were 1 or 2 bits smaller than
if we had allowed special $q$'s to be only on one side; the norms were
accordingly a bit smaller.

The general strategy used for a given special $q$ is classical: among a
vast quantity of candidates, we mark those that are divisible by primes
up to a given {\em sieving bound} using a sieve {\em \`a la} Eratosthenes;
then the most promising candidates are further scrutinized using the ECM
method, trying to extract primes up to the smoothness bound $B$. The
criterion for selecting those promising candidates is best expressed as
the number of times the smoothness bound is allowed for the remaining
part of the norms once the sieved primes have been removed. This is
usually referred to as the {\em number of large primes} allowed on a given
side.
\medskip

For the 1024-bit computation, we used the following parameters. On the
rational side, we sieved all the prime special $q$ in the 150M--1.50G
range (that is, with $1.5\cdot
10^8<q<1.5\cdot 10^9$), and on the algebraic side, we sieved special-$q$ prime ideals in
the range 150M--1.56G. The difference between the two is due to the fact that we
used two different clusters for this step, and when we stopped
sieving, one was slightly ahead of the other.

For each special $q$, the number of $(a,b)$-pairs considered was about
$2^{31}$. This number includes the pairs where both $a$ and $b$ are even,
but almost no time is spent on those, since they cannot yield a valid
relation.

All primes on both sides up to 150M were extracted using sieving, and the
remaining primes up to the smoothness bound $B = 2^{31}$ were extracted
using ECM. On the side where the special $q$ was placed, we allowed 2 large
primes, while $3$ large primes were allowed on the other side.
\medskip

This relation collection step can be parallelized almost infinitely
with no overhead since each special $q$ is handled completely
separately from the others. We used a variety of computational
resources for the sieving, and in general took advantage of
hyperthreading and in addition oversubscribed our virtual cores with
multiple threads.  Aggregating reported CPU time for virtual cores
over all of the machine types we used, we spent $5.08 \cdot 10^9$ CPU
seconds, or 161 CPU years sieving the rational side, and $5.03 \cdot
10^9$ CPU seconds, or 159 CPU years sieving the algebraic side.  In
order to obtain a more systematic estimate of the CPU effort dedicated
to sieving without these confounding factors, we ran sampling
experiments on a machine with 2 Intel Xeon E5-2650 processors running
at 2.00 Ghz with 16 physical cores in total.  From these samples, we estimate
that sieving would have taken 15 years on this machine, or 240
core-years.  We spent about one month of calendar time on sieving.

The total number of collected relations was 520M relations: 274M from the
rational side and 246M from the algebraic side. Among them, 249M were
unique, involving 201M distinct prime ideals. After filtering these
relations, we obtained a matrix with 28M rows and columns, with 200
non-zero entries per row on average.

Before entering the linear algebra step, we calculated the dense
block of ``Schirokauer maps'', which are $q$-adic characters introduced
by
Schirokauer in~\cite{Schirokauer93}. These consist, for each matrix row,
in $3$ full-size integers modulo $q$ (the number $3$ is here the unit
rank of the number field defined by our polynomial~$f$).

\subsection{Linear algebra}

The linear algebra problem to be solved can be viewed in several ways.
One is to consider a square matrix of size $N\times N$, whose left-hand side
$M$ of size $N\times(N-r)$ is the matrix produced by the filtering task,
while the right block $S$ of size $N\times r$ is made of dense
Schirokauer maps.
Recent work~\cite{JouPie16} has
coined the term ``nearly sparse'' for such matrices. We seek a non-trivial
element of the right nullspace of the square matrix $(M\|S)$.\footnote{%
    The integer factorization case, in contrast, has $q=2$, and 
    requires an element of the \emph{left} nullspace. The latter fact
    allows for a two-stage algorithm selecting first many solutions to
    $xM=0$, which can then be recombined to satisfy
    $x(M\|S)=0$. No such approach works for the right
    nullspace.} This approach has the drawback that an
iterative linear algebra algorithm based on the matrix $(M\|S)$ is
hampered by the weight of the block $S$, which contributes to each
matrix-times-vector product.

\subsubsection{Shirokauer maps serve as initialization vectors}
An alternative method, originally proposed by
Coppersmith~\cite[§8]{Coppersmith94} alongside the introduction of
the block Wiedemann algorithm, is to use this algorithm to constructively
write a zero element in the sum of the column spaces of $M$ and $S$. In
this case, the iterative algorithm is run on the square matrix
$M_0=(M\|0)$, which is of considerably lower weight than
$(M\|S)$.  More precisely, the block Wiedemann algorithm, with
two \emph{blocking factors} which are integers $m$ and $n$, achieves this
by proceeding through the following steps. The blocking factor $n$ is
chosen so that $n\geq r$, and we let $D(t)$ be the diagonal $n\times n$
matrix with coefficients $t$ ($r$ times) and $1$ ($n-r$ times).
\begin{description}
    \item[Initialization.] Pick blocks of \emph{projection vectors}
        $x\in\bF_q^{N\times m}$ and \emph{starting vectors}
        $y\in\bF_q^{N\times n}$. The block $x$ is typically chosen of
        very low weight, while we set $y=(S\|R)$, with $R$ a
        random block in $\bF_q^{N\times
        (n-r)}$.
    \item[Sequence.] Compute the sequence of matrices
        $a_i={}^t x M_0^i y$, for $0\leq i<L$, with $L=\left\lceil
        N/m\right\rceil + \left\lceil N/n\right\rceil + \left\lceil
        m/n+n/m\right\rceil$.
    \item[Linear generator.] Let $A(t)=\sum_i a_it^i$.  Let
        $A'(t)=A(t)D(t)\mathrel{\text{div}} t$.
        Compute an $n\times n$ matrix of
        polynomials $F(t)$ such that $A'(t)F(t)$ is a matrix of
        polynomials of degree less than $\deg F$, plus terms of
        degree above $L$ (see~\cite{Coppersmith94,Thome02b}). We
        typically have $\deg F\approx N/n$.
    \item[Solution.]
        Consider one column of degree $d$ of $F(t)$. Write the
        corresponding column of $D(t)F(t)$ as $c t^{d+1} + f_0t^d +
        \cdots + f_d$ with $c,f_i\in \bF_q^{n\times 1}$.  With high
        probability, we have $c\not=0$ and $w=(S\|0)c + M_0
        \sum_{i\geq0} M_0^i y f_i=0$. Rewrite that as $Mu+Sv=0$, where
        $u$ and $v$ are:
        \begin{align*}
            u&=\text{first $N-r$ coefficients of}\ \sum_{i\geq0} 
                                M_0^i y f_i,\\
            v&=\text{first $r$ coefficients of $c$}
        \end{align*}
        This readily provides a solution to the problem.
\end{description}


\subsubsection{Solving the linear system with $(1+o(1))N$ SpMVs}
The most expensive steps above are the \emph{sequence} and
\emph{solution} steps. The dominating operation is the sparse
matrix-times-vector operation (SpMV), which multiplies $M_0$ by a column
vector in $\bF_q^{N\times 1}$. It is easy to see that the sequence
step can be run as $n$ independent computations, each requiring $L$ SpMV
operations (therefore $nL=(1+n/m)N$ in total): matrices $a_i$ are
computed piecewise, column by column. Once all these computations are
completed, the fragments of the matrices $a_i$ need to be collated to a
single place in order to run the linear generator step.

It is tempting to regard the solution step in a directly similar way (as
was done, e.g., in the context of the recent 768-bit DLP
computation~\cite{DLP768}). However, as was pointed out very early on by
Kaltofen~\cite[Step C3, p. 785, and corollary to Theorem 7]{Kaltofen95}
yet seems to have been overlooked since, one should proceed differently.
Assume that some of the vectors $M_0^iy$ from the sequence step have been
kept as regular checkpoints (an obvious choice is $M_0^{Kj}y$ for some
well chosen checkpoint period $K$). For an arbitrary $j$, we compute
$\sum_{i=0}^{i=K-1}M_0^{i} M_0^{Kj}y f_{Kj+i}$ with a Horner evaluation
scheme which costs $K$ SpMV operations only. These expressions together
add up to $u$, and can be computed independently (using as many
independent tasks as $K$ allows). This adds up to $\deg F\approx N/n$
SpMV operations.

In total, this evaluation strategy yields a cost of $(1+n/m+1/n)N$ SpMV
operations (see~\cite[Theorem 7]{Kaltofen95}), which can be freely
parallelized $n$-fold for the sequence step, and possibly much more for
the solution step. It is important to note that as blocking factors $m$
and $n$ grow with $m\gg n$, this brings the total cost close to $N$ SpMV operations, a
count which to our knowledge beats all other exact sparse linear algebra
algorithms. The only limiting point to that is the linear generator step,
whose cost depends roughly linearly on $(m+n)$. Thanks to the use of
asymptotically fast algorithms~\cite{BeLa94,Thome02b,GiLe14}, this step
takes comparatively little time.

\subsubsection{Linear algebra for 1024-bit SNFS DLP}

%
%
%
%
%
The matrix $M_0$ had 28 million rows and columns, and 200 non-zero
coefficients per row on average. We used the linear algebra code in
CADO-NFS~\cite{cado-nfs-dev}.  We chose blocking factors $m=24$, $n=12$.
Consequently, a total of 44 million SpMV operations were needed. We ran
these in two computing facilities in the respective research labs of the
authors, with a roughly even split. For the sequence step, each of the 12
independent computations used between 4 and 8 nodes, each with up to 44
physical cores. The nodes we used were interconnected with various
fabrics, including Mellanox 56 Gbps Infiniband FDR and 40 Gbps Cisco UCS Interconnects.  The total time
for the sequence step was about 123 core-years. The linear generator step
was run on 36 nodes, and cost 13 core-years. The solution step was split
in $48=\frac{\deg F=2400000}{K=50000}$ independent tasks. Each used a
fairly small number of nodes (typically one or two), which allowed us to
minimize the communication cost induced by the parallelization. Despite
this, each iteration was 33\% more expensive than the ones for the
sequence step, because of the extra cost of the term $M_0^{Kj}y f_{Kj+i}$
which is to be added at each step. The total time for the solution step
was 9 core-years, which brings the total linear algebra cost for this
computation below 150 core-years.  In total we spent about one month of calendar time on linear algebra.
Table~\ref{tab:bwc-timings} in Appendix~\ref{appendix:bwc-timings} gives more details of the iteration times for the different machine architectures present in our clusters.

After this step and propagating the knowledge to relations that were
eliminated during the filtering step we obtained the logarithm of 198M
elements, or $94.5\%$ of the prime ideals less than $2^{31}$.

\subsection{Individual logarithms}

In our scenario where a malicious party has generated a trapdoored
prime with a goal of breaking many discrete logarithms in the
corresponding group, it is interesting to give details on the
individual logarithm step, which is often just quickly mentioned as an
``easy'' step (with the notable exception of~\cite{Logjam} where it is
at the heart of the attack).

\newcommand{\Binit}{B_{\mathrm{init}}}

From now on, we denote by $z$ the element of the group $G$ for
which we want the discrete logarithm (modulo $q$). The database of
discrete logarithms computed thus far is with respect to an a priori
unknown generator. In order to obtain the logarithm of $z$ with respect
to a generator specified by the protocol being attacked, it is typical
that two individual logarithm queries are necessary. This aspect will
not be discussed further.

The individual
logarithm step can itself be decomposed in two sub-steps:
\begin{description}
    \item[Initialization] Find an exponent $e$ such that
        $z'=z^e\equiv u/v \mod p$, where $u$ and $v$ are $\Binit$-smooth numbers
        of size about half of the size of $p$. Note that $\Binit$ has to
        be much larger than $B$ to get a reasonable smoothness
        probability.
    \item[Descent] For every factor of $u$ or $v$ that is larger than the
        smoothness bound, treat it as a special-$q$ to rewrite its
        discrete logarithm in terms of smaller elements, and continue
        recursively until it is rewritten in terms of elements of known
        logarithm.
\end{description}

We emphasize that the Initialization step does not use the 
polynomials selected for the NFS computation, and therefore, it does not take
advantage of the SNFS nature of the prime $p$. For the Descent step, on
the other hand, this makes heavy use of the polynomials, and here knowing the
hidden polynomials corresponding to $p$ helps significantly.

\subsubsection{Asymptotic complexities}
In terms of asymptotic complexity, the Initialization step is more costly
than the Descent step, and in a theoretical analysis, the bound $\Binit$ is
chosen in order to minimize the expected time of the Initialization step
only. The early-abort analysis of Barbulescu~\cite[Chapter 4]{BarbulescuPhD}
gives $B_i=L_p(2/3, 0.811)$, for a running time of
$L_p(1/3, 1.232)$.

For the Descent step, the complexity analysis can be found in two different
flavours in the literature: either we use polynomials of degree 1 (the
polynomial $a-bx$ corresponding to an $(a,b)$-pair) like
in~\cite{Semaev02}, or polynomials of
possibly higher degrees, depending on where we are in the descent
tree similarly to~\cite{JC:EngGauTho11}.

Using higher degree polynomials, we get a complexity of $L_p(1/3,
0.763)$, where the last steps of the descent are the most costly.
Sticking with polynomials of degree $1$, the first steps become more
difficult and the complexity is $L_p(1/3, 1.117)$. Both are lower than
the complexity of the Initialization step.

We give further details on the complexity analysis of individual logarithms in Appendix~\ref{appendix:individual}.

\subsubsection{Practical approach}
This theoretical behaviour gives only a vague indication of the
situation for our practical setting.
For the initialization step, we follow the general idea of
Joux-Lercier~\cite{JoLe03}.  For a random integer $e$, we compute $z'\equiv
z^e \mod p$, and take two consecutive values in the extended Euclidean
algorithm to compute $u_0$, $v_0$, $u_1$, $v_1$ of size about $\sqrt{p}$
such that $z'\equiv \frac{u_0}{v_0} \equiv \frac{u_1}{v_1} \mod p$. We
then look for two integers $a$ and $b$ such that $u = au_0 + bu_1$ and
$v=av_0 + bv_1$ are both smooth. Since this is an unlikely event, and
testing for smoothness is very costly, we do a 3-step filtering strategy.

First, we sieve on the set of small $(a,b)$-pairs, to detect pairs for
which the corresponding $u$ and $v$ are divisible by many small primes.
For this, we re-use the sieving code that helps collecting relations.
After this first step, we keep only $(a,b)$-pairs for which the remaining
unfactored part is less than a given threshold on each side.

Then, many ECM factorizations are run on each remaining cofactor;
this is tuned so that we expect most of the prime factors up to a
given bit size to be extracted. After this step, we again keep only the
candidates for which the remaining unfactored parts are smaller than
another threshold.
The cofactors of the surviving pairs are then fully factored using MPQS.

At each stage, if a prime factor larger than the smoothness bound $B$
is found, we naturally abort the computation.

This practical strategy keeps the general spirit of filters used in the
theoretical analysis of Barbulescu which relies only on ECM, but we found
that combined with sieving and MPQS, it is much faster.

\subsubsection{Parameters for the 1024-bit SNFS computation}
For the 1024-bit computation, we used a bound $\Binit=135$ bits
for this Initialization step. After applying the Joux-Lercier trick,
we first sieved to extract primes up to
$2^{31}$, and we kept candidates for which the unfactored parts are both
less than 365 bits. Then we used GMP-ECM with 600 curves and
$B_1=500,000$, hoping to remove most of the prime factors of 100 bits and
less. After this second step, we kept only the candidates for which the
unfactored parts are both less than 260 bits.
\medskip

For the Descent step, we used only polynomials of degree 1. Polynomials
of degree 2 do not seem to yield smaller norms even for the largest
135-bit primes to be descended. The depth of the recursion was rarely
more than 7. A typical example of a sequence of degrees encountered while
following the tree from the top to a leave is
$$ 135 \rightarrow 90 \rightarrow 65 \rightarrow 42 \rightarrow 33 \rightarrow 31,$$
but this is of course different if the ideals along the path are on the
rational or the algebraic side.
\medskip

Both the Initialization and the Descent steps can be heavily
parallelized. The expected CPU-time for computing an individual logarithm
is a few days on a typical core, distributed more or less equally between
the two steps. Using parallelization, we managed to get a result in 80
minutes of wall-clock time: the initialization took around 20 minutes parallelized 
across 500 cores, and the descent took 60 minutes parallelized across 352 cores.
(We used 44 cores for each large special $q$ to be descended.)
\medskip

As an example, we computed the discrete logarithm of
$$ z = \lceil \pi 10^{307}\rceil = 3141592653\cdots 7245871,$$
taking $2$ as a generator. More precisely, we are talking about their
images in the subfield of order $q$ obtained by raising $2$ and $z$ to
the power $(p-1)/q$. We obtained:
$$ \log z / \log 2 \equiv
409559101360774669359443808489645704082239513256 \mod q,$$
which can be easily checked to be the correct answer.

\section{Discussion}
\label{sec:discussion}

\subsection{Comparison with GNFS DLP for various sizes}

The recently reported GNFS 768-bit discrete logarithm
computation~\cite{DLP768} took about 5000 core-years. It is tempting to
directly compare this number to the 400 core-years that we spent in our
experiments. As a rule of thumb, one would expect the 768-bit GNFS to be about a factor of 10 more difficult than a 1024-bit SNFS computation, and this appears to hold in the numbers we report. However, we note that
first, the software used in both experiments are different (the CADO-NFS
sieving implementation is slower than Kleinjung's), and second, the GNFS-768
computation was done with a safe prime, while we used a DSA prime, thus
saving a factor of 6 in the linear algebra running time.

It is possible to get another hint for the comparison by considering the
typical sizes of the norms in both contexts. For GNFS-768, they appear to
be roughly 20 bits larger (in total) than for SNFS-1024. Taking all
correcting factors into account, like the $\alpha$-values, the
(un-)balance of the norms, and the special-$q$, this translates to roughly
a factor of 8 in the smoothness probability, thus more or less confirming
the ratio of running times observed in practice.
\medskip

Asymptotically, the difference of complexities between GNFS and SNFS
(namely the factor $2^{1/3}$ in the exponent) means that we would expect
to obtain similar running times when SNFS is run on an input that is twice the size of
the one given to GNFS. However, key sizes relevant to current practice and
practical experiments are still too small for these asymptotic bounds to be accurate.

To get concrete estimates for these smaller key sizes, we can compare the size of the 
norms and estimate that an 1152-bit SNFS computation would correspond to the same 
amount of time as a GNFS-768. For an SNFS of 2048 bits, the equivalent
would be around a GNFS of 1340 bits. And finally, for an SNFS of 4096
bits, the equivalent would be around a GNFS of 2500 bits. Of course, for
such large sizes these are more educated guesses than precise estimates.

\subsection{Non-hidden SNFS primes in real use}
\label{sec:nonhidden}
We have found multiple implementations using non-hidden SNFS primes in
the real world.  150 hosts used the 512-bit prime $2^{512}-38117$ for
export-grade Diffie-Hellman key exchange in a full IPv4 HTTPS scan
performed by Adrian et al.~\cite{Logjam} in March 2015.  Performing
the full NFS discrete log computation for this prime took about
215 minutes on 1288 cores, with 8 minutes spent on the sieving stage, 
145 minutes spent on linear algebra, and the remaining time spent filtering relations and reconstructing logarithms. In September 2016, 134 hosts were observed still using this prime.

We also found 170 hosts using the 1024-bit prime $2^{1024}-1093337$ for
non-export TLS Diffie-Hellman key exchange in scans performed by Adrian
et al. In September 2016, 106 hosts were still using this prime.  We
estimate that performing a SNFS-DL computation for this prime would
require about 3 times the amount of effort for the sieving step as the
1024-bit SNFS computation that we performed. This difference is mostly
due to the $\alpha$-value of the $f$-polynomial that can not easily be
made small. The linear algebra step will suffer at the very least a
7-fold slowdown. Indeed, since this prime is safe, the linear algebra
must be performed modulo $(p-1)/2$, which is more expensive than the
160-bit linear algebra we used for a DSA prime in our computation.
Furthermore, since the smoothness probabilities are worse, we expect also
the matrix to be a bit larger, and the linear algebra step cost to grow
accordingly.

The LibTomCrypt library~\cite{libtomcrypt}, which is widely distributed and provides public domain implementations of a number of cryptographic algorithms, includes several hard-coded choices for
Diffie-Hellman groups ranging in size from 768 to 4096 bits.  Each of
the primes has a special form amenable to the SNFS.  The 768-bit strength (actually a 784-bit
prime) is $2^{784} - 2^{28} + 1027679$.  We performed a SNFS discrete log for
this prime. On around a thousand cores, 
we spent 10 calendar days sieving and 13 calendar days on 
linear algebra. The justification for the
special-form primes appears to be the diminished radix form suggested
by Lim and Lee~\cite{limlee}, which they suggest for decreasing the
cost of modular reduction.  We examined the TLS and SSH scan datasets collected
by Adrian et al.~\cite{Logjam} and did not find these primes in use for either protocol.

We also carried out a perfunctory search for poorly hidden SNFS primes
among public key datasets, based on the rather straightforward strategy
in~§\ref{sec:unveil}, hoping for monic $g$, and $f$ such that $2\leq d\leq 9$ and $|f_d|\leq 1024$.
We carried out this search for the 11,658 distinct SSH group exchange
primes, 68,126 distinct TLS ephemeral Diffie-Hellman primes, and
2,038,232 distinct El Gamal and DSA primes from a dump of the PGP public
key database.  This search rediscovered the special-form TLS primes
described above, but did not find any other poorly hidden primes
susceptible to SNFS.  We cannot rule out the existence of trapdoored primes
using this method, but if hidden SNFS primes are in use the designers
must have followed Gordon's advice.

 
\subsection{Lessons}

It is well known among the cryptographic community that 1024-bit
primes are insufficient for cryptosystems based on the hardness of
discrete logarithms.  Such primes should have been removed from use
years ago.  NIST recommended transitioning away from 1024-bit key
sizes for DSA, RSA, and Diffie-Hellman in 2010~\cite{sp800-131a}.
Unfortunately, such key sizes remain in wide use in practice.  Our
results are yet another reminder of the risk, and we show this
dramatically in the case of primes which lack verifiable randomness.
The discrete logarithm computation for our backdoored prime was only
feasible because of the 1024-bit size.

The asymptotic running time estimates suggest that a SNFS-based
trapdoor for a 2048-bit key would be roughly equivalent to a GNFS
computation for a 1340-bit key.  We estimate that such a computation
is about 16 million times harder than the 1024-bit computation that we
performed, or about $6.4 \cdot 10^9$ core-years.  Such a computation
is likely still out of range of even the most sophisticated
adversaries in the near future, but is well below the security guarantees that a
2048-bit key should provide.  Since 2048-bit keys are likely to remain
in wide usage for many years, standardized primes should be published
together with their seeds.

In the 1990s, key sizes of interest were largely limited to 512 or
1024 bits, for which a SNFS computation was already known to be
feasible in the near future.  Both from this perspective, and from our
more modern one, dismissing the risk of trapdoored primes in real
usage appears to have been a mistake, as the apparent difficulties
encountered by the trapdoor designer in 1992 turn out to be easily
circumvented.  A more conservative design decision for FIPS 186 would
have required mandatory seed publication instead of making it
optional.  As a result, there are opaque, standardized 1024-bit and
2048-bit primes in wide use today that cannot be properly verified.

\ifanon
\else
\section*{Acknowledgements}  
We are grateful to Paul Zimmermann for numerous discussions all along
this work.
Rafi Rubin performed invaluable system administration for the University of Pennsylvania cluster.  Shaanan Cohney and Luke Valenta contributed to sieving for the 784-bit SNFS-DL computation. 
Part of the experiments presented in this paper were carried out using the
Grid'5000 testbed, supported by a scientific interest group hosted by
Inria and including CNRS, RENATER and several Universities as well as
other organizations.
We are grateful to Cisco for donating the Cisco UCS hardware that makes up most of the University of Pennsylvania cluster.   Ian Goldberg donated time on the CrySP RIPPLE Facility at the University of Waterloo and Daniel J. Bernstein donated time on the Saber cluster at TU Eindhoven for the 784-bit SNFS-DL computation.  This work was supported by the U.S. National Science foundation under grants CNS-1513671, CNS-1505799, and CNS-1408734, and a gift from Cisco.
\fi

\bibliographystyle{abbrv}
\bibliography{biblio}

\appendix

\section{Complexity analysis of individual logarithms}
\label{appendix:individual}

The complexity analysis of individual logarithms is not well detailed
in the literature, in particular in the SNFS case. For convenience we
summarize the results in this appendix. As usual in the NFS context,
the claimed complexities are not rigorously proven and
rely on heuristics.

The notation is the same as in the main body of the paper: $p$ is a
prime and we have to compute the
discrete logarithm of an element $z$ in a prime order subgroup of
$\bF_p^*$. We are given $f$ and $g$ a pair of polynomials that have been
used for an NFS computation so that the (virtual) logarithms of all the
ideals of norm less than a bound $B$ should have been
pre-computed. The bound $B$, the degrees of $f$ and $g$, and the sizes
of their coefficients depend on the General vs Special NFS variant.

We recall the classical corollary of the Canfield-Erd\H os-Pomerance
theorem that expresses smoothness probabilities in terms of the
$L$-notation:
\begin{theorem}
    Let $a,b,u,v$ be real numbers such that $a>b>0$ and $u,v>0$. As
    $x\rightarrow\infty$, the
    proportion of integers below $L_x(a,u)$ that are
    $L_x(b,v)$-smooth is
    $$L_x\Big(a-b,\frac uv(a-b)+o(1)\Big)^{-1}.$$
\end{theorem}

\subsection{Initialization of the descent}

This step consists of first ``smoothing'' $z$ in order to bootstrap
the subsequent descent step. We choose a random integer $e$, compute the element
$z'\equiv z^e \mod p$, and test it for smoothness. Many
elements are tested until one is found to be $\Binit$-smooth. The best known
algorithm for smoothness testing is the ECM algorithm: it extracts (with
high probability) all primes up to a bound $K$ in time
$L_K(1/2,\sqrt{2}+o(1))$. The dependence in the size of the integer from
which we extract primes is polynomial, so we omit it: in our
context this type of factor ends up being hidden in the $o(1)$ in the
exponents.

From this estimate, one can derive that if we want to allow a running
time in $L_p(1/3, \cdot)$, then $\Binit$ can only be as large as $L_p(2/3,
\cdot)$; otherwise, testing the smoothness would be too costly. At the
same time, the probability that $z'$ is $\Binit$-smooth drives the number
of attempts and puts additional constraints. It is remarkable that it
also imposes a smoothness bound $\Binit$ in $L_p(2/3,\cdot)$ to get an
$L_p(1/3,\cdot)$ number of attempts. Following
Commeine-Semaev~\cite{CoSe06}, if we set $\Binit = L_p(2/3, c)$, one can
show that the expected running time for the basic algorithm for the
initialization step is in $L_p(1/3, \frac{1}{3c}+2\sqrt{c/3}+o(1))$, which
is minimal for $c=1/\sqrt[3]{3}$, yielding a complexity of $L_p(1/3,
\sqrt[3]{3}+o(1)) \approx L_p(1/3, 1.442)$.

Inspired by the early abort strategy that Pomerance~\cite{Pomerance82} had
developed in the context of the quadratic sieve,
Barbulescu~\cite{BarbulescuPhD} has shown that this complexity can be
reduced. The idea is to start the smoothness test with a bound smaller
than the target $\Binit$ smoothness bound: this allows one to extract the
smallest factors. Then, we make a decision based on the size of the
remaining unfactored part: if it is too large, the
probability that this will yield a $\Binit$-smooth number is too small
and we start again with another random exponent $e$. In other words,
instead of testing immediately for $\Binit$-smoothness, we first run a filter, 
with cheaper ECM parameters, that allows us to select promising
candidates for which the full test is run. Analyzing this technique and
optimizing all the involved parameters is not a simple task; according
to~\cite{BarbulescuPhD}, we obtain a final complexity of $\approx
L_p(1/3, 1.296)$ for a smoothness bound $\Binit=L_p(1/3, 0.771)$.

This is not the end of the story: instead of just one filter, one can add
more. The analysis becomes even more involved, but this improves
again on the complexity. Numerical experiments indicate that performance
does not improve beyond 6 filters, and for 6 filters, the
final complexity given in~\cite{BarbulescuPhD} is summarized in the following fact: 
\smallskip

\noindent{\bf Fact.} The initialization step of the descent can be done
in time $L_p(1/3, 1.232)$ with
a smoothness bound $\Binit=L_p(1/3, 0.811)$.
\medskip

Finally, we mention that writing $z'\equiv\frac{u}{v}\mod p$ for $u$ and
$v$ that are about half the size of $p$, and testing them for smoothness does not
change the asymptotic complexities, but it yields a huge practical
improvement, especially when combined with sieving as in
Joux-Lercier~\cite{JoLe03}.

On the other hand, when neither $f$ nor $g$ are linear polynomials, the
smoothing test has to be done in one of the number fields, and then, in
this context, using half-size elements is necessary to get the
appropriate complexity; we refer to~\cite[Section 8.4.3]{BarbulescuPhD}
for details about this.

\subsection{Descent step}

\newcommand{\gq}{\mathfrak{q}}

After the initialization step, the discrete logarithm of $z$ can be expressed
in terms of the virtual logarithms of a few ideals of degree~$1$ in one
of the number fields associated to $f$ or $g$. Those whose norm is less
than the smoothness bound $B$ that was used in the sieving and linear
algebra steps are assumed to be already known. Since $B=L_p(1/3,\cdot)$
while $\Binit=L_p(2/3,\cdot)$, we expect to have a handful of prime
ideals whose logarithms are not known. These are the ones that will be
subject to this descent step. We do the analysis for one ideal of maximal
size $\Binit$; since there are only polynomially many of them, doing all
of them will contribute only to the $o(1)$ in the final exponent.

Let $\gq$ be an ideal of norm $q=L_p(\alpha, c)$, where
$\alpha\in[\frac13,\frac23]$. We consider the lattice of
polynomials $\varphi(x) = a_0+a_1x+\cdots+a_{k-1}x^{k-1}$ that, after
being mapped to a principal ideal in the number field where $\gq$
belongs, become divisible by $\gq$. For $k=2$, this would correspond to
the $(a,b)$-pairs corresponding to $\gq$ seen as a special-$q$, but we
allow larger degrees. Since we are going to allow a search that takes a
time $T$ in $L_p(1/3, \cdot)$ for handling $\gq$, the $a_i$'s can be
bounded by $(qT)^{1/k} = L_p(\alpha,c/k)L_p(1/3, \cdot)$.

Let us analyze first the case where $\alpha>1/3$ so that the second
factor can be neglected. The product of the norms is given by
$$\begin{array}{rcl}
    \Res(f(x), \varphi(x))\Res(g(x),\varphi(x)) &\approx &
\|\varphi\|^{\deg f+\deg g} \|f\|^{k-1}\|g\|^{k-1} \\
    & \approx & L_p(\alpha, c/k(\deg f+\deg g)) (\|f\|\,\|g\|)^{k-1}.\\
\end{array} $$
Let us write $\deg f + \deg g = \delta(\log p/\log\log p)^{1/3}$, so
that we can cover all the variants. Then $\|f\|\,\|g\|$ is $L_p(2/3,
2/\delta)$ in the case of GNFS and $L_p(2/3, 1/\delta)$ in the case of
SNFS.  Finally, the product of the norms is
$$L_p\left(\alpha+1/3,\frac{c\delta}{k}\right)L_p\left(2/3,{\scriptstyle\left\{\genfrac{}{}{0pt}{}{1\text{
    for SNFS}}{2\text{
    for GNFS}}\right\}}\frac{k-1}\delta\right).$$
Here there are two strategies: we can fix $k=2$, so that the second
factor does not contribute to the complexity, or we can let $k$ grow in order to
balance the two factors.

\subsubsection{Descending with higher degree polynomials.} The best value
for $k$ is proportional to $(\log p/\log\log p)^{\alpha/2-1/6}$ (we
deliberately omit to analyze the proportionality ratio).
In
that case, the product of the norms takes the form
$$ L_p(\alpha/2+1/2, \cdot),$$
so that, since we allow a time $L_p(1/3,\cdot)$, we can expect to find
an element that is $L_p(\alpha/2+1/6, \cdot)$-smooth. The smoothness test
implies multiplying the cost by $L_p(\alpha/4+1/12, \cdot)$, which is
bounded by $L_p(1/4,\cdot)$ since $\alpha \leq 2/3$, and therefore does
not contribute to the final complexity. As a consequence, as long as $\alpha$ is
more than $\frac13$, it is possible to descend a $\gq$ whose norm is in
$L_p(\alpha, \cdot)$ in prime ideals of norms at most $L_p(\alpha/2+1/2,
\cdot)$, in time bounded by $L_p(1/3, \cdot)$. We can choose
the exponent constant smaller than the other steps of the descent so that
these first steps become negligible. This is true whether we are dealing
with GNFS or SNFS.

As we get close to $\alpha=\frac13$, the value of $k$ tends to a
constant. We postpone the corresponding analysis.

\subsubsection{Descending with degree-1 polynomials.}
In the case where we force $k=2$, the product of the norms is dominated
by the first factor and we get $L_p(\alpha+\frac13, c\delta/2)$.
Let us try to descend $\gq$ in prime ideals of norms slightly smaller
than the norm $q$ of $\gq$, namely we target $L_p(\alpha, c\lambda)$,
for some value $\lambda$ that we hope to be strictly less than $1$.
The probability of the product of the norms being $q^\lambda$-smooth is
then in $L_p(\frac13, \frac{\delta}{6\lambda}+o(1))^{-1}$. The cost of
smoothness testing with ECM is in $L_p(\frac{\alpha}{2}, \cdot)$, which
is negligible as soon as $\alpha<2/3$. Hence, the cost of the descent
with degree-1 polynomials is dominated by the case $\alpha=2/3$, which
we will now focus on. In this limiting case, the cost of ECM is
$L_{L_p(2/3, c\lambda)}(1/2, \sqrt{2}+o(1)) = L_p(1/3,
2\sqrt{c\lambda/3}+o(1))$, so that the time to descend $\gq$ in prime ideals
of norms bounded by $q^\lambda$ is in $L_p(1/3, \frac{\delta}{6\lambda} +
2\sqrt{c\lambda/3}+o(1))$. This is minimized for
$\lambda=\sqrt[3]{\delta^2/12c}$ and yields a running time of $L_p(1/3,
(3c\delta/2)^{1/3}+o(1))$. In the case of GNFS, we have $\delta=3^{1/3}$,
while it is $\delta=(3/2)^{1/3}$ for SNFS. We fix $\lambda$ so that we
minimize the time when dealing with the largest $\gq$ coming out from the
initialization step, namely for $q=L_p(2/3, 0.811)$; this value $c=0.811$
gives $\lambda=0.598$ in the case of GNFS, and $\lambda=0.513$ in the
case of SNFS. Both are less than 1, which means that the descent process
indeed descends. Finally, we obtain the following:
\smallskip

\noindent{\bf Fact.} If we use degree-1 polynomials, the cost of the
first stages of the descent is $L_p(1/3, 1.206)$ for GNFS and $L_p(1/3,
1.117)$ for SNFS.

\subsubsection{Last steps of the descent.} We now deal with the final
steps of the descent where $\gq=L_p(1/3, c)$, with $c$ larger than the
constant involved in the smoothness bound $B=L_p(1/3, \cdot)$, which
depends on whether we are in the GNFS or SNFS case. In this setting, there is
no gain in considering $k>2$, so we keep $k=2$. The factor that was
neglected when evaluating the size of the $a_i$'s is no longer
negligible, so we start again, and assume that we are going to spend
time $T=L_p(1/3, \tau+o(1))$. This propagates into the formulae and gives
a bound $L_p(1/3, (\tau+c)/2)$ for the $a_i$'s, which in turn gives
$$ L_p(2/3, (\tau+c)\delta/2)\|f\|\, \|g\|$$
for the product of the norms. Let us denote $B=L_p(1/3, \beta)$ the
smoothness bound used for sieving and linear algebra, and write
$c=\beta+\varepsilon$, where $\varepsilon>0$. We omit the details, but it
can be checked that if we allow time $L_p(1/3, \beta)$, we can
descend $\gq$ in prime ideals of norms at most $L_p(1/3,
\beta+\frac{\varepsilon}{4})$. This analysis is valid both for GNFS and
SNFS, even though the values of $\beta$ and $\delta$ are different for
these two cases. This is no surprise that this is the cost of finding
one relation in the sieving step, since when $\gq$ is just above the
smoothness bound, descending involves essentially the same procedure as
what we do during sieving with special-$q$ that are marginally smaller.
We obtain therefore:
\smallskip

\noindent{\bf Fact.} The cost of the last stages of the descent is
$L_p(1/3, 0.961)$ for GNFS and $L_p(1/3, 0.763)$ for SNFS.
\medskip

In this analysis, we have not studied the transition between the two
modes where we decrease the value $\alpha$ or the value $c$ when
descending an ideal of size $L_p(\alpha, c)$. This technicality is dealt
with in~\cite{JC:EngGauTho11} in the context of the Function Field Sieve, but it
applies {\em mutatis mutandis} to our NFS situation.
\medskip

In the following table, we summarize the exponent constants in the
$L_p(1/3,\cdot)$ complexities of the various steps of the descent, for
GNFS and SNFS, allowing or not sieving with higher degree polynomials:
\begin{center}
    \begin{tabular*}{\textwidth}{c @{\ \ } c @{\extracolsep{\fill}} c c @{\extracolsep{\fill}} c}
    \toprule
    &  & \multicolumn{3}{c}{Descent step} \\
        \cmidrule(r){3-5}
    &  Initialization step& \multicolumn{2}{c}{Large $\gq$} & Small $\gq$ \\
    &                     & Sieving $\deg=1$ & Sieving higher deg &  \\
    \midrule
    GNFS & $1.232$ & $1.206$ & $o(1)$ & $0.961$ \\
    SNFS & $1.232$ & $1.117$ & $o(1)$ & $0.763$ \\
    \bottomrule
\end{tabular*}
\end{center}

\section{Block Wiedemann algorithm timings}
\label{appendix:bwc-timings}

%
%

\begin{table}[h]
  \begin{center}
        \begin{tabular}{c@{~~}@{~}c@{~~}c@{~~}c@{~~}c@{~~}c@{~~}c@{~}}
    \noalign{\global\belowrulesep=3pt\global\aboverulesep=3pt}
          \toprule
Location & Nodes & CPU Type & Clock Speed & Cores & RAM & Interconnect\\
\midrule
\ifanon Institute 1\else UPenn\fi & 20 & 2$\times$Xeon E5-2699v4 & 2.2 -- 2.8GHz & 44 & 512GB & eth40g\\
& 8 & 2$\times$Xeon E5-2680v3 &2.5 -- 2.9GHz & 24 & 512GB & eth40g\\
& 6 & 2$\times$Xeon E5-2699v3 & {2.3 -- 2.8GHz} & 36 & 128GB & eth10g\\
\ifanon Institute 2 \else Nancy\fi & 48 & 2$\times$Xeon E5-2650v1 & {2.0 -- 2.4GHz} & 16 & 64GB & ib56g\\
\bottomrule
\end{tabular}
    \end{center}
\caption{We ran both sieving and linear algebra on various clusters of different configurations.  For the CPU clock speed, we give both nominal and turbo speeds.}

\label{tab:cluster-config}
\end{table}

\begin{table}[h]
    \begin{center}
        \begin{tabular*}{\textwidth}{c c c @{\extracolsep{\fill}} c c c}
          \toprule
          CPU Type & Interconnect & Nodes/Job & \multicolumn{3}{c} {Seconds per iteration}\\
\cmidrule(r){4-6}
& & & \hfil Sequence & Solution & Communication \\
\midrule
Xeon E5-2699v4 & eth40g & 1 &  & 2.42 & 0.12\\
&  &  4 & 0.41 &  & 0.17\\
&  &  8 & 0.19 &  & 0.17\\
&  &  12 & 0.13 &  & 0.14\\
&  &  16 & 0.10 & 0.21 & 0.13\\
Xeon E5-2680v3 & eth40g & 2 &  & 2.24 & 0.30\\
&  &  8 & 0.35 &  & 0.15\\
Xeon E5-2699v3 & eth10g & 6 & 0.36 &  & 0.33\\
Xeon E5-2650v1 & ib56g & 2 &  & 3.7 & 0.19\\
&  & 8 & 0.60 &  & 0.10\\
\bottomrule
\end{tabular*}
    \end{center}
\caption{ Timings for the Block Wiedemann algorithm as run on the
  various clusters for the 1024-bit SNFS Discrete Log computation.
  Table~\ref{tab:cluster-config} gives details on the node
  configurations.}

\label{tab:bwc-timings}
\end{table}

\end{document}